\documentstyle[aps,epsfig]{revtex}
\newcommand{\be}{\begin{equation}}
\newcommand{\ee}{\end{equation}}
\newcommand{\bea}{\begin{eqnarray}}
\newcommand{\eea}{\end{eqnarray}}
\newcommand{\st}{{\scriptscriptstyle T}}
\newcommand{\xbj}{x}
\newcommand{\zh}{z}

\def\F{{D^q_1(z,z^2 k_\st^2)}}
\def\Fs{{H_1^{\perp q}(z,z^2 k_\st^2)}}

\begin{document}
\tighten
\thispagestyle{empty}
\title{
\begin{flushright}
\begin{minipage}{3 cm}
\small
hep-ph/9701330\\ 
NIKHEF 97-002
\\ VUTH 97-1
\end{minipage}
\end{flushright}
Probing transverse quark polarization via azimuthal asymmetries 
in leptoproduction}
\author {{\bf A.M. Kotzinian}\footnote
{on leave from
Yerevan Physics Institute, AM-375036 Yerevan, Armenia
and JINR, RU-141980 Dubna, Russia } \\
{\it Universit\"at Mainz, D-55099 Mainz, Germany}\\
E-mail: Aram.Kotzinian@cern.ch\\
and\\
{\bf P.J. Mulders}\\
{\it Department of Physics and Astronomy, Free University of Amsterdam}\\
{\it and National Institute for Nuclear Physics and High-Energy Physics}\\
{\it P.O. Box 41882, NL-1009 DB Amsterdam, the Netherlands}\\
E-mail: pietm@nikhef.nl}

\maketitle

\begin{abstract}
We consider the leading order result for polarized leptoproduction,
putting emphasis on transverse momentum dependent effects appearing
in azimuthal asymmetries.
Measurements of weighted cross sections enable extraction of the 
distribution of transversely polarized quarks. We focus on the
distribution in a longitudinally polarized hadron and estimate the
expected asymmetries in leptoproduction.
\end{abstract}

  
\section{Introduction}

For the study of azimuthal distributions of hadrons produced in deep inelastic
lepton-hadron scattering, the transverse momenta of quarks with respect
to the hadron in the quark distribution functions (DF's) and in the 
quark fragmentation functions (FF's) play an important role, even in leading 
order in $1/Q$ where $Q^2 = -q^2$ is the momentum transfer squared. 
Omitting details that can be found in \cite{tm95,mt96} and restricting
ourselves to leading order in $1/Q$,
the six DF's needed to describe the quark density matrix
in a nucleon depend on $x$ and $p_\st$ which parametrize (the relevant
part of) the quark momentum $p$, in a nucleon with momentum $P$, 
$p = x\,P + p_\st$. The subscript $T$ refers to transverse with respect
to the momenta of target hadron ($P$) and produced hadron ($P_h$).
For a polarized nucleon the spin vector is written 
as $S = \lambda\,P/M + S_\st$, satisfying $\lambda^2 - S_\st^2$ = 1.
Then, the probability ${\cal P}^q_N(x,p_\st^2)$, 
the longitudinal spin distribution $\lambda^{(in)\,q}(x,p_\st)$, 
and the transverse spin distributions $s_T^{(in)\,q}(x,p_\st)$, 
of the quark in a polarized nucleon are given by
\bea
&&{\cal P}^q_N(x,p_\st) = f^q_1(x,p_{\st}^2), \\
&&{\cal P}^q_N(x,p_\st)\, \lambda^{(in)\,q}(x,p_\st) =
\lambda\,g^q_{1L}(x,p_\st^2)
- \frac{p_\st\cdot S_\st}{M}\,g^q_{1T}(x,p_\st^2),\\
&&{\cal P}^q_N(x,p_\st) \, s_\st^{(in)\,q}(x,p_\st) \,
= S_\st\,h^q_{1T}(x,p_{\st}^2) 
+ \frac{p_\st}{M}\left[ \lambda h_{1L}^{\perp\,q}(x,p_\st^2) - 
\frac{p_\st\cdot S_\st}{ M}\,h_{1T}^{\perp\,q}(x,p_\st^2)\right].
\eea
In inclusive processes one only encounters $p_\st$-integrated results.
Integrating the lefthandside over $p_\st$ one finds on the
righthandside $x$-dependent distribution functions,
\bea
&&\int d^2p_\st \,{\cal P}^q_N(x,p_\st) = f^q_1(x), \\
&&\int d^2p_\st \,{\cal P}^q_N(x,p_\st)\, \lambda^{(in)q}(x,p_\st) =
\lambda\,g^q_1(x), \\
&&\int d^2p_\st \,{\cal P}^q_N(x,p_\st) \, s_\st^{(in)q}(x,p_\st) =
S_\st\,\left( h^q_{1T}(x) + h_{1T}^{\perp (1)q}(x)\right)
\equiv S_\st\,h^q_1(x),
\eea
where we have straightforwardly $p_\st$-integrated functions and 
$(-p_\st^2/2M^2)$-weighted functions indicated with superscript $(1)$.
\bea
f(x)&=&\int d^2p_\st \,f(x,p_\st^2) , \\
f^{(1)}(x)&=&\int d^2p_\st \,\left(\frac{-p_\st^2}{2M^2}\right)\,f(x,p_\st^2). 
\eea
The function $f^q_1(x)$ is the familiar quark distribution function, also
often denoted as $q(x)$, $g^q_1(x)$ and $h^q_1(x)$ are the longitudinal and
transverse spin quark distribution functions, also often denoted as 
$\Delta q(x)$ and $\Delta_Tq(x)$.
Two other $(-p_\st^2/M^2)$-weighted functions appear in the single 
$(p_\st/M)$-weighted results for polarized quarks,
\bea
&&
\int d^2p_\st\,\frac{p_\st}{M}\,{\cal P}^q_N(x,p_\st)\, 
\lambda^{(in)q}(x,p_\st) =
S_\st\,g_{1T}^{(1)q}(x), \\ &&
\int d^2p_\st \,{\cal P}^q_N(x,p_\st) 
\,\frac{p_\st\cdot s_\st^{(in)q}(x,p_\st)}{M} =
\lambda\,h_{1L}^{\perp (1)q}(x).
\eea

In the analysis of the current jet in hadroproduction one encounters
at leading order fragmentation functions depending on $z$ and 
$P_{h\perp}$ -- the hadron transverse momenta with respect to the
quark momenta, which describe the decay of a quark with momentum $k$ into a 
hadron with momentum $P_h$ = $z\,k + P_{h\perp}$. This is equivalent
to a quark with momentum $k$ = $P_h/z + k_\st$ producing 
a hadron with momentum $P_h$ provided its transverse momentum is
given by $k_\st$ = $-P_{h\perp}/z$.
For the case that no polarization in the final state is measured one
has quark fragmentation functions, defined via the quark decay function
\be
{\cal D}^q_h(z,k_\st) = D^q_1(z,P_{h\perp}^2=z^2 k_\st^2)
+ \frac{\epsilon_\st^{ij}k_{\st i}\,s_{\st j}^{(out)q}}{M_h}
\,H_1^{\perp q}(z,z^2 k_\st^2),
\ee
where $s_\st^{(out)q}$ is the transverse polarization of the fragmenting 
quark. The second fragmentation function allowing the possibility 
of a correlation between the produced hadron transverse momentum and
the transverse polarization of fragmenting quark is nonzero
because of non-applicability of time reversal invariance in a decay
process~\cite{Gas66,RKR71,HHK83,mt96}. This specific function was first
discussed by Collins~\cite{col}.
Upon $k_\st$-integration of the lefthandside one
finds the following nonvanishing combinations
\bea
&&\int d^2P_{h\perp} {\cal D}^q_h(z,k_\st) =
\int d^2 P_{h\perp} \,D^q_1(z,P_{h\perp}^2) 
= z^2\int d^2k_\st\,\F 
\equiv D^q_1(z), \\
&&\int d^2P_{h\perp} \,{\cal D}^q_h(x,k_\st) 
\,\frac{\epsilon_T^{ij} k_{\st i}\,s_{\st j}^{(out)q}}{M_h}
= z^2\int d^2 k_\st \,\left(\frac{-k_\st^2}{2M_h^2}\right)
\,H_1^{\perp q}(z,z^2k_\st^2)
\equiv H_1^{\perp (1)q}(z).
\eea
The function $D^q_1(z)$ is the familiar fragmentation function, normalized
through the momentum sum rule $\sum_h \int dz\,zD_1^{q\rightarrow h}(z) = 1$. 

One of the reasons to consider transverse momentum dependent distribution
and fragmentation function is their appearance in measurements of 
azimuthal asymmetries in Drell-Yan scattering, 1-particle inclusive 
leptoproduction, or in jet analysis in $e^+e^-$ annihilation. On the
theoretical side there is the relation of transverse momentum dependent
functions and higher twist functions as discussed in ref.~\cite{mt96}.
We mention particularly the relations with the twist-three quark
distributions $g^q_T$ and $h^q_L$,
\bea
g^q_T(x) = g^q_1(x) + \frac{d}{dx}\,g_{1T}^{(1)q},
\label{relation1}
\\
h^q_L(x) = h^q_1(x) - \frac{d}{dx}\,h_{1L}^{\perp(1)q}.
\label{relation2}
\eea
The first distribution appears in the structure function $g^q_2$ 
measured in inclusive deep inelastic scattering, while the latter
appear for instance in Drell-Yan asymmetries~\cite{jj}.
The first relation was discussed
earlier in a slightly different framework in ref.~\cite{bkl84}.

\section{The polarized semi-inclusive cross section}

The cross section for 1-particle inclusive deep inelastic scattering 
is given by
\be
\frac{d\sigma^{(\ell +N\rightarrow \ell^\prime+h+N)}}
{d\xbj\,dy\,d\zh\,d\phi^\ell\,d^2P_{h\perp}} =
\frac{\alpha^2}{2Q^4}\,\frac{y}{2\zh}\,L_{\mu \nu}\, 2M{\cal W}^{\mu \nu},
\ee
where the scaling variables are defined as $\xbj$ = $Q^2/2P\cdot q$,
$y$ = $P\cdot q/P\cdot l$ ($l$ is a momentum of incoming lepton) 
and $\zh$ = $P\cdot P_h/P\cdot q$. We have not bothered to introduce
different notations for the momentum fractions $x$ and $z$ used in 
the previous section and the scaling variables as they will be identified 
in the leading order calculation.
The transverse space (e.g. $P_{h\perp}$) is defined with respect to the
momenta $P$ and $q$. The azimuthal angles are angles in the transverse
space giving the orientation of the lepton plane ($\phi^\ell$) and the
orientation of the hadron plane ($\phi^\ell_h$ = $\phi_h - \phi^\ell$) 
or spin vector ($\phi^\ell_s$ = $\phi_s - \phi^\ell$)
with respect to the lepton plane. 
The angles all are defined around the z-axis defined by 
the momenta $P$ and $q$.  The quantity $L_{\mu\nu}$
is the well-known lepton tensor, while the hadronic tensor is
given by
\bea
2M{\cal W}_{\mu\nu}( q; P S; P_h ) & = &
\int \frac{d^3 P_X}{(2\pi)^3 2P_X^0}
\,\delta^4 (q + P - P_X - P_h)
\nonumber \\
&&\qquad \qquad \times
\langle P S |J_\mu (0)|P_X; P_h \rangle
\langle P_X; P_h |J_\nu (0)|P S \rangle.
\label{hadrten}
\eea
At leading order the calculation only involves the DF's and FF's 
discussed in the previous section, and each quark (and antiquark) contributes
\bea
2M\,{\cal W}_{\mu\nu} & = & 2\zh \sum_q e_q^2\int d^2k_\st\,d^2p_\st
\,\delta^2\left(p_\st - k_\st - \frac{P_{h\perp}}{\zh}\right)
\,f^q_1(\xbj,p_\st^2)\,D^q_1(\zh,\zh^2 k_\st^2)
\nonumber \\
&& \qquad\qquad \times \Bigl[ - g_\st^{\mu\nu}
+ i\lambda^{(in)}\,\epsilon_\st^{\mu\nu}
+ s_\st^{(in)\,\{\mu}s_{\st}^{(out)\,\nu\}}
-s_\st^{(in)}\cdot s_{\st}^{(out)}\,g_\st^{\mu\nu} 
\Bigr]
\eea
Performing the contraction this leads to the cross section as shown first in
\cite{ak95} with the following terms
\begin{equation}
\frac{d\sigma^{\ell+N\rightarrow \ell^\prime+h+X}}
{dxdyd\phi^\ell d\zh d^2P_{h\perp}}=\frac{\alpha^2}{Q^2 y} \sum_q e_q^2 
\sigma_q,
\end{equation}
where
\begin{eqnarray}
\sigma_q &=& \int d^2p_\st\,d^2k_\st
\,\delta^2\left(p_\st - k_\st - \frac{P_{h\perp}}{\zh}\right)
\nonumber \\ 
&& \qquad \mbox{} \times \Biggl\{
\left[1+(1-y)^2\right]\,f^q_1(\xbj,p_\st^2)D^q_1(\zh,\zh^2k_\st^2)
\nonumber \\ 
&& \qquad\qquad \mbox{} + \lambda_e \lambda \,y(2-y)
\,g^q_1(\xbj,p_\st^2)D^q_1(\zh,\zh^2k_\st^2)
\nonumber \\ 
&&\qquad\qquad \mbox{} -\lambda_e \,y(2-y) \,\frac{(p_\st\cdot S_\st)}{M}
\,g^q_{1T}(\xbj,p_\st^2)D^q_1(\zh,\zh^2k_\st^2)
\nonumber \\ 
&&\qquad\qquad \mbox{} - 2(1-y)
\,\frac{ k_\st^1 S_\st^2+k_\st^2 S_\st^1}{M_h} 
\,h^q_{1T}(\xbj,p_\st^2) H_1^{\perp q}(\zh,\zh^2 k_\st^2)
\nonumber \\ 
&&\qquad\qquad \mbox{} - \lambda \,2(1-y) 
\,\frac{k_\st^1 p_\st^2+k_\st^2 p_\st^1}{MM_h}
\,h_{1L}^{\perp q}(\xbj,p_\st^2) H_1^{\perp q}(\zh,\zh^2 k_\st^2)
\nonumber \\
&&\qquad\qquad \mbox{} + 2(1-y) 
\,\frac{p_\st \cdot S_\st}{M}\,\frac{p_\st^1 k_\st^2+p_\st^2 k_\st^1}{MM_h}
\,h_{1T}^{\perp q}(\xbj,k^2)H_1^{\perp q}(\zh,\zh^2 k_\st^2)
\Biggr\} 
\end{eqnarray}

Now let us consider the differential cross section for one quark flavour, 
$\sigma_q$ , integrated with different weights depending on the final
hadron transverse momenta and the direction of the nucleon transverse
polarization with respect to virtual photon direction, 
$w_i(P_{h\perp},\hat S_\st)$:
\be
I_i=\int d^2P_{h\perp}\, w_i(P_{h\perp},\hat S_\st)\,\sigma_q
\ee

\begin{enumerate}
\item
$w_1(P_{h\perp},\hat S_\st)=1$.

Taking into account that  
$I_i = \int d^2k_\st\, d^2p_\st 
\,w_i\left(z(p_\st-k_\st), \hat S_\st\right)\{...\}$
and that odd powers of $k_\st^1, k_\st^2$ and $p_\st^1, p_\st^2$
give zero contribution to $I_i$, we find cross sections involving
the transverse momentum-integrated distribution and fragmentation
functions
\be
I_1=[1+(1-y)^2]\,f_1(x) D_1(\zh)+\lambda_e \lambda\,y(2-y)\,g^q_1(x)D^q_1(\zh).
\ee

\item
$w_2(P_{h\perp},\hat S_\st)=(-P_{h\perp}\cdot \hat S_\st/\zh)
=(\vert P_{h\perp}\vert/\zh)\,\cos (\phi_h^\ell - \phi_s^\ell)
=(k_\st - p_\st) \cdot \hat S_\st$.

The surviving terms upon integration are
\be
I_2 =
\lambda_e M\vert S_\st\vert\,y(2-y)\,g_{1T}^{(1) q}(\xbj)D^q_1(\zh)
- M_h\vert S_\st\vert \,2(1-y) \,h^q_1(\xbj)H_1^{\perp (1) q}(\zh)
\,\sin 2\phi_s^\ell,
\ee
Note that the first term is proportional to the lepton polarization, 
while the second is not.
One can also see that integrating over $\phi_s^\ell$ (which is equivalent to
integration over $\phi^\ell$) the second term $\propto \sin 2\phi_s^\ell$ 
in the above result vanishes and we get the result of our 
previous paper (Ref.~\cite{km95}).

\item
$w_3(P_{h\perp},\hat S_\st)
=(P_{h\perp}^1\hat S_\st^2+P_{h\perp}^2\hat S_\st^1)/\zh
=(\vert P_{h\perp}\vert/\zh)\,\sin(\phi_s^\ell+\phi_h^\ell)$.

The surviving contributions are
\be
I_3 =
\lambda_e M\vert S_\st\vert\,y(2-y)\,g_{1T}^{(1) q}(\xbj)D^q_1(\zh)
\,\sin 2\phi_s^\ell
- M_h\,\vert S_\st\vert\,2(1-y)\,h^q_1(\xbj) H_1^{\perp (1)q}(\zh).
\ee
As in the previous case the second term is independent of the lepton 
polarization and appears due to the Collins single spin asymmetry.

\item
$w_4(P_{h\perp},\hat S_\st)=
(P_{h\perp}^1 P_{h\perp}^2/2\zh^2)
=(\vert P_{h\perp}\vert^2/4\zh^2)\,\sin 2\phi_h^\ell$.

The surviving contribution is
\be
I_4=MM_h\,\lambda\,2(1-y)
\,h_{1L}^{\perp (1)q}(\xbj)H_1^{\perp(1)q}(\zh).
\ee

\item
One can use another $w(P_{h\perp},\hat S_\st)$ to get separate 
contributions from $h^q_{1T}(x,p_\st^2) $ and $h_{1T}^{\perp q}(x,p_\st^2)$.
For example  $w_5(P_{h\perp},\hat S_\st)
\propto P_{h\perp}^1 P_{h\perp}^2 (P_{h\perp}\cdot \hat S_\st)$
will give another combination of $h^q_{1T}(x,p_\st^2) $ and
$h_{1T}^{\perp q}(x,p_\st^2)$, containing higher (in
transverse momentum space) moments of distribution and fragmentation
functions. But to separate the contribution of $h_{1L}^{\perp q}(x,p_\st^2)$ 
the weight factor $w_4$ suffices.
\end{enumerate}

\section{Approximations}

In ref.~\cite{km95} we investigated $g_{1T}^{(1)q}$ by employing the 
relation with $g_T$ (Eq.~\ref{relation1})
and the approximation of the latter by the 
Wandzura-Wilczek part determined by the polarized quark distribution
function $g^q_1$. This led to
\be
g_{1T}^{(1)q}(x) \simeq x\int_x^1 dy\,\frac{g^q_1(y)}{y},
\ee
Here we want to follow a similar route and use the relation between
$h_{1L}^{\perp(1)q}$ and $h^q_L$ (Eq.~\ref{relation2})
and the separation of the latter in
interaction-independent and interaction-dependent parts\cite{mt96},
\be
h^q_L(x) = \frac{2}{x}\,h_{1L}^{\perp(1)q} + \frac{m}{Mx}\,g^q_1 + \tilde h^q_L.
\ee
Omitting the interaction-dependent term $\tilde h^q_L$ and the quark mass
term and using Eq.~\ref{relation2} one gets the differential equation
\be
2\frac{d}{dx}h_{1L}^{\perp(1)q}(x)-\frac{h_{1L}^{\perp(1)q}(x)}{x}
= h^q_1(x),
\ee
leading with the requirement $h_{1L}^{\perp(1)q}(1)=0$ to
\be
h_{1L}^{\perp(1)q}(x) \simeq -\frac{1}{2}\,x \int_x^1 dy\,\frac{h^q_1(y)}{y}.
\ee
Next one can use for an order of magnitude estimate 
$h^q_1(x)\simeq g_1(x)$,  which is valid for example in the bag
model~\cite{jj}. Thus, in this approximation 
\be
h_{1L}^{\perp(1)q}(x) \simeq -\frac{1}{2}\,x \int_x^1 d y\frac{g^q_1(y)}{y}.
\ee
We will use for numerical estimation the values obtained using
parametrization of DF's from Ref.~\cite{bbs}.

\begin{figure}[!h]
\begin{center}
\mbox{ \psfig{file=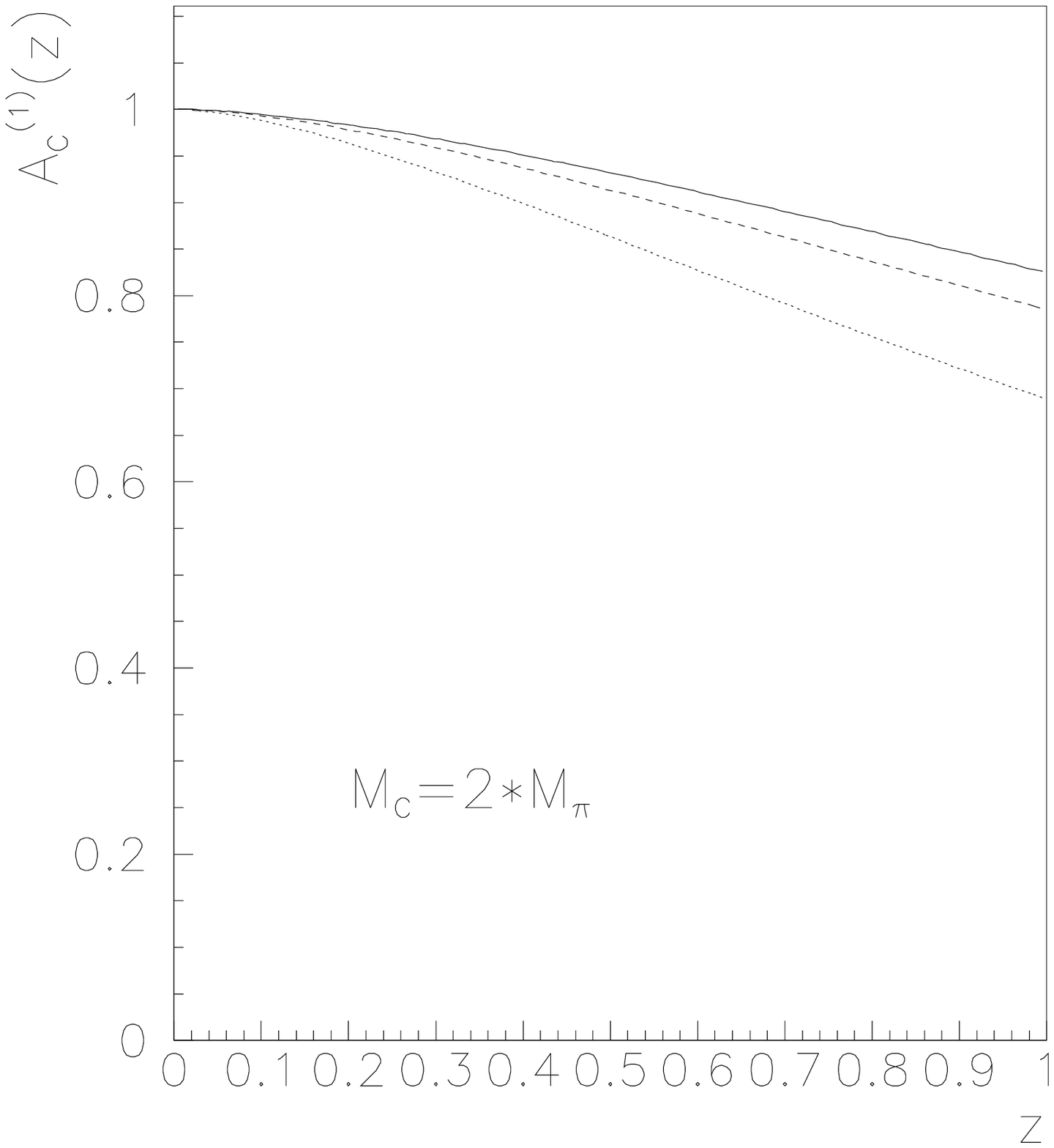,height=7.cm,width=0.45\hsize} }
\mbox{ \psfig{file=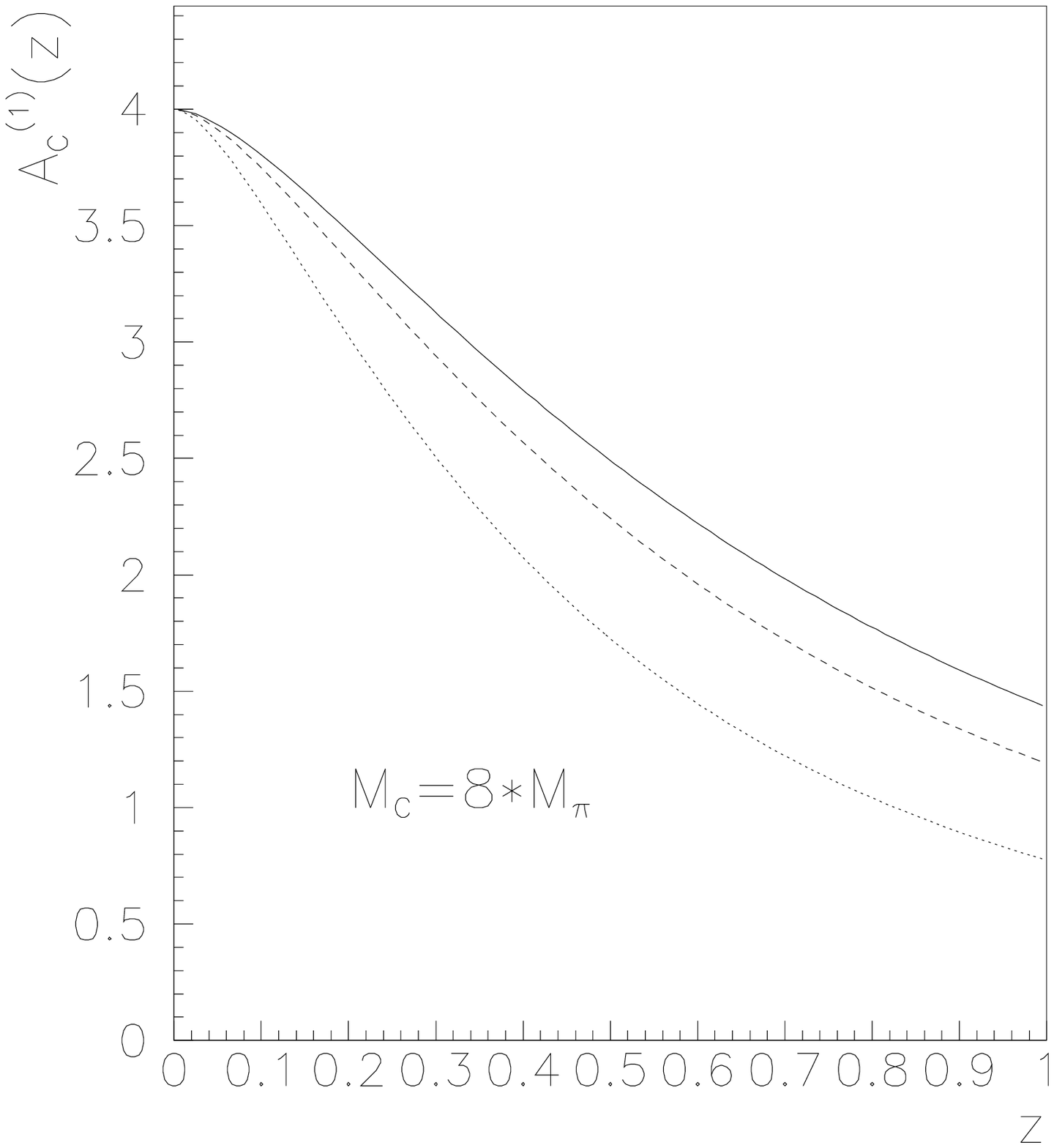,height=7.cm,width=0.45\hsize} }
\end{center}
\caption{$H_1^{\perp(1)u}(z)/D^u_1(z)$ for different values of $M_C$;
solid line -- $\langle P_{h\perp}^2 \rangle=0.98$~(GeV/c)$^2$,
dashed line -- $\langle P_{h\perp}^2 \rangle=0.70$~(GeV/c)$^2$
and dotted  line -- $\langle P_{h\perp}^2 \rangle=0.36$~(GeV/c)$^2$.  }
\label{fig:fsf}
\end{figure}

Next, let us turn to the estimate of $H_1^{\perp(1)q}(z)$. 
Collins~\cite{col} suggested the following parametrization for the 
analyzing power in transversely polarized quark fragmentation
\be
A_C(z,k_T) \equiv \frac{\vert k_T\vert\Fs}{M_h\F}
=\frac{M_C\,\vert k_T\vert}{M_C^2+\vert k_T^2\vert },
\ee
where $M_C \simeq 0.3\div 1.0$~GeV is a typical hadronic mass. This
parametrization exhibits the kinematic zero when $k_T=0$, the leading
twist asymmetry when $k_T={\cal{O}}(M)$, and the higher twist asymmetry
when $k_T\gg M$. Now, assuming
a Gaussian parametrization for the unpolarized fragmentation function 
\be 
\F=D^q_1(z)\,\frac{R^2}{\pi\,z^2}\,\exp(-R^2 k_T^2),
\ee
one obtain 
\be
H_1^{\perp (1)q}(z)=D^q_1(z)\frac{M_C}{2M_h}\left(1-M_C^2R^2
\int_{0}^{\infty}dx\,\frac{\exp(-x)}{x+M_C^2R^2}\right).
\ee
Note, that $R^2=z^2/\langle P_{h\perp}^2 \rangle$ where 
$\langle P_{h\perp}^2 \rangle$
is a hadron mean-square momentum in the quark fragmentation.
According to different analyses \cite{exp} 
$\langle P_{h\perp}^2 \rangle \simeq 0.36 \div 0.98$ (GeV/c)$^2$.
Fig.~\ref{fig:fsf} represents the ratio 
$A_C^{(1)}(z)\equiv H_1^{\perp (1)u}(z)/D^u_1(z)$ 
for different values of $\langle P_{h\perp}^2 \rangle$ and $M_C$.

\section{Numerical results}

We will consider production of $\pi ^{+}$-mesons on the proton
target. In order to get an order of magnitude estimate we note 
that the cross section is given predominantly by scattering on 
the $u$-quark.
Consider the target longitudinal spin asymmetry defined as
\be
A(x,y,z:\lambda) \equiv 
\frac{\int d\phi^l \int d^2P_{h\perp} \frac{\vert P_{h\perp}^2\vert}{4z^2M_h^2}
\sin 2\phi_h^\ell \left(d\sigma^+-d\sigma^-\right)}
{\int d \phi^l \int d^2P_{h\perp} \left(d\sigma^+ + d\sigma^-\right)},
\ee
where $+ (-)$ denotes target positive (negative) longitudinal polarization.
Using $I_1$ and $I_4$ we see that for both polarized and unpolarized
lepton this asymmetry is given by
\be
A(x,y,z;\lambda) =-\lambda\,\frac{2(1-y)}{1+(1-y)^2}
\,\frac{h_{1L}^{\perp(1)u}(x)H_1^{\perp(1)u}(z)}{f^u_1(x) D^u_1(z)}.
\ee
Note, that this expression is valid for unpolarized as well as for
polarized lepton beam.

For polarized leptons one can consider also the asymmetry defined as
\be
A_1(x,y,z) \equiv 
\frac{\int d \phi^l \int d^2P_{h\perp} \frac{\vert P_{h\perp}^2\vert}{4z^2M_h^2}
\sin 2\phi_h^\ell \left(d\sigma^{++}-d\sigma^{+-}\right)}
{\int d \phi^l \int d^2P_{h\perp} \left(d\sigma^{++}-d\sigma^{+-}\right)},
\ee  
where the first (second) superscript of $d\sigma$ denotes lepton (target) 
polarization, leading to
\be
A_1(x,y,z) = \frac{2(1-y)}{y(2-y)}
\,\frac{h_{1L}^{\perp(1)u}(x)H_1^{\perp(1)u}(z)}{g^u_1(x) D^u_1(z)}.
\ee

With the approximation for $h_{1L}^{\perp (1)u}$ and using
the ratio $x\int_x^1dy\frac{g^u_1(y)}{y}/f^u_1(x)$ as calculated
in~\cite{km95} (see Fig. 3 there), which reaches a maximal value 
of 0.08 at $x\simeq 0.5$, we obtain for small $y$-values 
and moderate $z$-values 
$A(x\simeq 0.5, y\simeq 0.1, z \simeq 0.3;
\lambda) \simeq (0.04 \div 0.12) \,\lambda$. 

The asymmetry $A_1$ is related to $A$,
\be
A_1(x,y,z) =
\frac{1+(1-y)^2}{\lambda \,y(2-y)}\,\frac{f^u_1(x)}
{g^u_1(x)}A(x,y,z;\lambda).
\ee
The ratio $g^u_1(x)/f^u_1(x)$ is presented in Fig. 1 of ref.~\cite{km95}
and leads to
$A_1 (x\simeq 0.2\div 0.5, y\simeq 0.1, z\simeq 0.3) \simeq 0.4 \div 2.8$.

Finally, let us consider the following weighted target transverse-spin
asymmetry:
\be
A_T(x,y,z;\vert S_T\vert) \equiv 
\frac{\int d \phi^l \int d^2P_{h\perp}\, \frac{\vert P_{h\perp}\vert}{zM_h}
\sin(\phi_s^\ell + \phi_h^\ell)
\,\left(d\sigma^{\uparrow}-d\sigma^{\downarrow}\right)}
{\int d \phi^l \int d^2 P_{h\perp} (d\sigma^{\uparrow}+d\sigma^{\downarrow})},
\ee
where $\uparrow (\downarrow)$ denotes target up (down) transverse polarization.
Using $I_1$ and $I_3$ we see that for both polarized and unpolarized
lepton this asymmetry is given by
\be
A_T(x,y,z,\vert S_T\vert) =
-\vert S_T\vert\,\frac{2(1-y)}{1+(1-y)^2}
\,\frac{h^u_1(x)H_1^{\perp(1)u}(z)}{f^u_1(x) D^u_1(z)}.
\ee
With the approximation $h^u_1(x) \simeq g^u_1(x)$ and using the ratio 
$g^u_1(x)/f^u_1(x)$ from ref.~\cite{km95} we see that asymmetry 
$A_T (x \simeq 0.2 \div 0.5, y \simeq 0.1,z \simeq 0.3;\vert S_T\vert)
\simeq -(0.4 \div 2.1)\vert S_T\vert$.
Thus, one sees that the $A_T$ asymmetry is an order of magnitude 
larger than $A$. Remember that both asymmetries arise due to the Collins
effect in transversely polarized quark fragmentation 
but in the second case this quark polarization is coming from 
the intrinsic transverse momentum.

In Ref.~\cite{col} an estimate for $H_1^{\perp q}/D^q_1$ has been given.
In Ref.~\cite{mt96} a full analysis of lepton-hadron scattering was presented,
including transverse momentum dependence in distribution and fragmentation
functions. The $k_\st$-moments of these functions are related to twist-3
functions. For the latter we consider only the 'interaction-independent'
part, which involves twist-2 functions. For the latter we finally assume
that the helicity and transverse spin distributions are identical. This
allows us to crudely estimate azimuthal asymmetries expected in a number
of observables. 
The transverse polarization of a quark in a longitudinally polarized nucleon 
arises due to intrinsic transverse momentum effects and is proportional to
$p_\st \, h_{1L}^{\perp\,q}$, which vanishes at $p_\st = 0$,
whereas in the transversely polarized nucleon it can be nonzero at $p_\st=0$.
This is the reason that the polarization azimuthal asymmetry for transversely
polarized nucleons ($A_T$) is much larger than for
longitudinally polarized nucleons ($A$).

A.K. is grateful to COMPASS Collaboration colleagues for valuable discussion.
The work of P.M. was supported by the foundation for Fundamental Research on
Matter (FOM) and the Dutch Organization for Scientific Research (NWO).


\begin{thebibliography}{99}
\bibitem{tm95} R.D.~Tangerman and P.J.~Mulders, Phys. Rev. {D 51}
(1995) 3357. 
\bibitem{mt96} P.J.~Mulders and R.D.~Tangerman, Nucl. Phys. {B 461} (1996) 197.
\bibitem{Gas66} 
S. Gasiorowicz, {\em Elementary Particle physics}, Wiley (New York), 1966.
\bibitem{RKR71}
A. de R\'ujula, J.M. Kaplan and E. de Rafael, 
Nucl. Phys. {B 35} (1971) 365. 
\bibitem{HHK83}
K. Hagiwara, K. Hikasa and N. Kai, Phys. Rev. {D 27} (1983) 84. 
\bibitem{col} J.~Collins, Nucl. Phys. {B 396} (1993) 161.
\bibitem{jj} R.L.~Jaffe and X.~Ji, Phys. Rev. Lett. {67} (1991)
552; Nucl. Phys. {B 375} (1992) 527.
\bibitem{bkl84} A.P. Bukhvostov, E.A. Kuraev and L.N. Lipatov, SP JETP
{60} (1984) 22.
\bibitem{ak95} A.~Kotzinian, Nucl. Phys. {B 441} (1995) 234.
\bibitem{km95} A.M.~Kotzinian and P.J.~Mulders, Phys. Rev. {D 54}
(1996) 1229.
\bibitem{bbs} S.J.~Brodsky, M.~Burkardt and I.~Shmidt, Nucl. Phys. 
{B 441} (1995) 197.
\bibitem{exp} EMC Collaboration, M.~Arneodo et al., Z. Phys. {C 34} 
(1987) 277; \\ 
EMC Collaboration, J. Ashman et al., Z. Phys. {C 52} (1991) 361; \\ 
E665 Collaboration, M.R.~Adams et al., Phys. Rev. {D 48} (1993) 5057.

\end{thebibliography}
\end{document}